\documentclass[journal]{IEEEtran}

\usepackage[tbtags]{amsmath}
\usepackage{amsfonts}
\usepackage{eurosym}
\usepackage{graphics}
\usepackage{graphicx}
\usepackage{array}
\usepackage{shortvrb}
\usepackage{epsf}
\usepackage{graphicx}
\usepackage{rotating}
\usepackage{float}
\usepackage{color}
\usepackage{multirow}
\usepackage{float}

\newcommand{\bfg}[1]{\mbox{\boldmath $#1$\unboldmath}}
\newcommand{\fraca}[2]{\displaystyle\frac{#1}{#2}}
\newcommand{\mm}[3]{\renewcommand{\arraystretch}{0.8}\begin{array}[t]{c}\mbox{#1}
\\ #2\end{array}\begin{array}[t]{c}#3\end{array}
\renewcommand{\arraystretch}{1}}
\def \R {{\rm I\kern -2.2pt R\hskip 1pt}}
\newtheorem{algorithm}{Algorithm}
\begin{document}
\title{Dynamic Robust Transmission Expansion Planning}

\author{Raquel~Garc\'{\i}a-Bertrand,~\IEEEmembership{Senior Member,~IEEE}
        and Roberto~M\'{\i}nguez
\thanks{
R. Garc\'{\i}a-Bertrand is partly supported by Junta de Comunidades de Castilla-La Mancha under project POII-2014-012-P; the Ministry of Science of Spain, under CICYT Project ENE2015-63879-R (MINECO/FEDER, UE); and the Universidad de Castilla-La Mancha under Grant GI20163388.
\newline \indent
R. Garc\'{\i}a-Bertrand is with the Department of Electrical Engineering, Universidad de Castilla-La
Mancha, Ciudad Real, Spain (e-mail: \mbox{Raquel.Garcia@uclm.es})
\newline \indent
R. M\'{\i}nguez is with Hidralab Ingeniería y Desarrollo, S.L., Spin-Off UCLM, Hydraulics Laboratory, Universidad
 de Castilla-La Mancha, Ciudad Real E-13071, Spain (e-mail: \mbox{roberto.minguez@hidralab.com}).}
}

\markboth{}%
{}

\maketitle

\begin{abstract}
Recent breakthroughs in Transmission Network Expansion Planning (TNEP) have demonstrated that the use of robust optimization, as opposed to stochastic programming methods, renders the expansion planning problem considering uncertainties computationally tractable for real systems. However, there is still a yet unresolved and challenging problem as regards the resolution of the dynamic TNEP problem (DTNEP), which considers the year-by-year representation of uncertainties and investment decisions in an integrated way. This problem has been considered to be a highly complex and computationally intractable problem, and most research related to this topic focuses on very small case studies or used heuristic methods
and has lead most studies about TNEP in the technical literature to take a wide spectrum of simplifying assumptions. In this paper an adaptive robust transmission network expansion planning formulation is proposed for keeping the full dynamic complexity of the problem. The method overcomes the problem size limitations and computational intractability associated with dynamic TNEP for realistic cases.
Numerical results from an illustrative example and the IEEE 118-bus system are presented and discussed, demonstrating the benefits of this dynamic TNEP approach with respect to classical methods.
\end{abstract}


\section*{Notation}
\renewcommand{\labelitemi}{}
The main notation used throughout this paper is stated below for quick
reference.
Other symbols are defined as needed throughout the paper.

{\bf Constants:}

\begin{description}

\item[$b_{k}$] Susceptance of line $k$ (S).

\item[$c^{\rm G}_{i}$] Generation cost for generator $i$ (\euro/MWh).

\item[$c^{\rm U}_{j}$] Load-shedding cost for consumer $j$ (\euro/MWh).

\item[$c_{k}$] Investment cost of building line $k$ (\euro).

\item[${\bfg D}^{(t)}$] Uncertainty set for time period $t$.

\item[$e^{(t)}_j$] Percentage of load shed by the $j$-th demand for year $t$.

\item[$f_{k}^{\rm max}$] Capacity of line $k$ (MW).

\item[$I$] Discount rate, i.e. the rate of return that could be earned on an investment in the financial markets with a similar risk.

\item[$n_y$] Number of study periods (years).

\item[$o(k)$] Sending-end bus of line $k$.

\item[$r(k)$] Receiving-end bus of line $k$.

\item[$\Pi$] Maximum budget for investment in transmission expansion (\euro).

\item[$\sigma$]  Weighting factor for obtaining annual operation and load-shedding costs (h).

\end{description}

{\bf Primal variables:}
\begin{description}
\item[${\bfg d}^{(t)}$] Vector of continuous variables representing the random or uncertain parameters for year $t$, i.e. generation capacities and loads (MW).

\item[$f^{(t)}_{k}$] Power flow through line $k$ for year $t$ (MW).

\item[$g^{(t)}_{i}$] Power produced by the $i$-th generating unit for year $t$ (MW).

\item[$p^{(t)}_{j}$] Power consumed by the $j$-th demand for year $t$ (MW).

\item[$r^{(t)}_{j}$] Load shed by the $j$-th demand for year $t$ (MW).

\item[$x_k^{(t)}$] Actual status ({\em existing} vs {\em no existing}) of line $k$ at the beginning of year $t$.

\item[$y_k^{(t)}$] Binary variable representing the construction of new line $k$ at the beginning of year $t$.

\item[$\theta^{(t)}_{n}$] Voltage angle at bus $n$ for year $t$ (radians).

\end{description}

{\bf Dual variables:} Note that dual variables are provided after the corresponding equalities or inequalities separated by a colon.\\

{\bf Indices and Sets:}
\begin{description}
%
%
%
\item[$\mathcal{D}^{(t)}$] Set of indices of demand for year $t$.

\item[$\mathcal{G}^{(t)}$] Set of indices of generating units for year $t$.

\item[$\mathcal{L}$] Set of all prospective and existing transmission lines at the beginning of time horizon considered.

\item[$\mathcal{L^{+}}$] Set of all prospective transmission lines.

\item[$\mathcal{N}$] Set of all networks buses.

\item[$n(i)$] Bus index where the $i$-th generating unit is located.

\item[$n(j)$] Bus index where the $j$-th demand is located.

\item[$\mathcal{T}$] Set of indices of different uncertainty sets.

\item[$\Psi_{n}^{\rm D}$] Set of indices of the demand located at bus $n$.

\item[$\Psi_{n}^{\rm G}$] Set of indices of the generating units located at bus $n$.

\end{description}

\section{Introduction}
\subsection{Motivation}
Transmission network expansion planning analyzes the issue of how to expand or reinforce an existing power transmission network to adequately service system loads over a given time horizon. This problem is challenging for several reasons \cite{LumbrerasR:16}:
\begin{enumerate}
    \item Transmission investments are capital intensive and have long useful lives (up to 40 years), which makes transmission investment decisions to have a long-standing impact on the power system as a whole.

    \item Vast amounts of new generation facilities, specially renewable, are expected to be built in the medium-term future. Effective transmission investments must integrate appropriately those new generation sources.

    \item It has been recognized that the uncertainties present in the problem, such as uncertainties associated with i) consumption, ii) renewable energy generation, such as with wind and solar power plants, and iii) equipment failure, constitute a considerable burden for its resolution.

    \item Energy production and use is interconnected with many other aspects of modern life, such as water consumption, use of goods and services, transportation, economic growth, land use, and population growth. Thus, changes in any of these variables might influence generation and/or load demands, and thereby affect transmission network expansion planning. Moreover, all these changes which affect transmission network expansion planning occur gradually and are subject to a high degree of uncertainty, therefore we have to adapt to those changes progressively, in such a way that transmission network expansion planning is less prone to inaccurate future prognosis of demand loads and power production.
\end{enumerate}

For the reasons given above, TNEP is by nature a multi-stage problem that entails planning a horizon of approximately 10 years, where the planner takes decisions at several time horizons which are reviewed every 2 years in the light of the revealed uncertainties \cite{DeDiosSC:07,RuizC:15}. However, the complexity of this dynamic nature has lead most studies about TNEP in the technical literature to take a wide spectrum of simplifying assumptions:
\begin{itemize}
   \item {\it Static approach:} Most research studies only consider one target year and planning and investment costs are considered annually (see for instance \cite{Garver:70,delaTorreCC:08,GarcesCGR:09,Jabr:13,SaumaO:06}).
  \item {\it Sequential static approach:} In this case, several target years are considered. It takes into account that any investments made will be available from their deployment date to the end of the planning horizon \cite{CosteiraT:11}, and it basically attempts to ensure tractability while keeping the model accurate enough. In this approach, the selected target years are treated in a separate and sequential way, i.e. TNEP problems associated with different years are solved sequentially assuming that the lines chosen to be constructed in a particular year are available for posterior periods. The final TNEP consists of the addition of those partial plans. The problem with this approach is that it loses the global perspective. The solution of a dynamic planning problem will not, in general, be the same as the collection of solutions associated with target year plans (see for instance \cite{PereiraPCO:85,BinatoOA:01,RochaS:11}).
  \item {\it Dynamic planning approach:} It keeps the full dynamic complexity of the problem. However, up to now, use of an integrated year-by-year representation of investment decisions has been considered to be a highly complex and computationally intractable problem, and most research related to this topic focuses on very small case studies or used heuristic methods \cite{EscobarGR:04,BragaS:05,Moeini-AghtaieAF:12}.
\end{itemize}

Static planning is the most reasonable approach when dealing with short time horizons where decisions are not going to be revisited. For longer time horizons, dynamic TNEP reproduce closely the reality of the problem, however, it cannot be implemented in real problems due to size limitations and computational intractability. The preferred method for those cases is the {\it sequential static approach}.

\subsection{Literature Review on TNEP Considering Uncertainties}
TNEP problems considering uncertainty have been dealt with using stochastic programming \cite{CarrionAA:07,LopezPQ:07,GarcesCGR:09,RohSW:09} and robust optimization \cite{WuCX:08,YuCW:11}.
Consideration of the effects of climate change in the generation of expansion planning problems has been put forward by \cite{LiCSF:14} by using a stochastic method with different future scenarios considered. However, stochastic programming formulations result in computationally intractable problems for real-size networks. In contrast, recent breakthroughs in robust TNEP problems \cite{Jabr:13,ChenWWHW:14,RuizC:15,MinguezG:15} proved that computational tractability for real-size systems is possible by using Adaptive Robust Optimization (ARO) frameworks \cite{ThieleTE:10,BertsimasLSZZ:13}. Besides, robust optimization is the recommended approach for the consideration of long-term uncertainties \cite{LumbrerasR:16}.

ARO materializes in a three-level formulation: i) the first level minimizes the cost of expansion (\cite{ChenWWHW:14} and also minimizes the maximum regret), the decision variables are those related to construction or expansion of lines, ii) the second level selects the least desirable outcome for the uncertain parameters maximizing the system's operational costs within the uncertainty set, the variables associated with this level are the uncertain generation capacities and demand, and iii) the third level selects the optimal decision variables to minimize operational costs for given values of first and second level variables. The main difference in methods that use  ARO make in TNEP is their way of  solving problems and how the uncertainty set is defined.

Specifically spreaking, \cite{Jabr:13} and \cite{MinguezG:15} merge the second and third levels into one single-level maximization problem using the third level dual. To deal with uncertain parameters and considering that they have to be equal to their upper or lower limits in the least desirable situation, binary variables are used. The limitation of this simplification is that the uncertainty budget must be an integer. However, this strategy does not belittle the benefits of robust optimization, but rather simplifies resolution of the problem substantially. Additionally, \cite{MinguezG:15} cuts the binary variables by half because the least desirable operational costs occur for generation capacities and demand loads below and above their nominal values \cite{RuizC:15}, respectively. Conversely, \cite{ChenWWHW:14} and \cite{RuizC:15} merge the second and third levels into one single-level maximization problem by using the Karush-Kuhn-Tucker (KKT)   third level conditions. Here, the number of constraints, continuous and binary variables of the subproblem increase with respect to the alternative approach. Note that in all approaches, authors deal with cardinality constrained uncertainty sets.

Once the third-level formulation is merged into a bi-level problem, \cite{Jabr:13} puts forward the Benders approach where the dual information from subproblems is used to construct additional Benders cuts. The main drawback with this method is the slow convergence typical with this type of decomposition algorithms \cite{ConejoCMG:06}, which made the author include additional linear constraints in order to improve convergence. Conversely, \cite{ChenWWHW:14,RuizC:15,MinguezG:15} apply a column-and-constraint generation method \cite{ZengZ:13} solely based on primal cuts. This  is computationally advantageous with respect to Benders decomposition and converges in a small number of iterations.

\subsection{Contribution}
The purpose of this paper is threefold:
\begin{enumerate}
    \item To extend the ARO formulation proposed by \cite{MinguezG:15} for the dynamic approach.
    \item To demonstrate how the dynamic model enables more optimal use of existing financial resources, rendering the solution more robust with respect to the initial selection of uncertainty sets and less prone to wrongful future prognosis of demand loads and generation capacities.
    \item To show that computational tractability for an integrated year-by-year representation of investment decisions ({\it dynamic approach}) is possible for realistic cases, ensuring the achievement of a global optimal solution.

\end{enumerate}

In summary, as a major contribution of this paper, we address a yet unresolved and challenging problem which is of utmost practical interest since it circumvents the simplifying assumptions typically adopted in the {\it static} and {\it sequential static} models available in the literature.

\subsection{Paper Structure}

The remainder of the paper is structured as follows. Section~\ref{Model} describes the dynamic adaptive robust formulation of the TNEP problem. In Section~\ref{s2}, the uncertainty set is defined and how it might change over the time horizon is shown. In Section ~\ref{s3}, the proposed solution approach is described.
Numerical results for an illustrative example and a realistic case study are given in Section~\ref{CaseStudy}. Finally, the paper is concluded in Section~\ref{Conclusions}.

\section{Dynamic Robust Transmission Network Expansion Planning Formulation}\label{Model}
 A detailed formulation of the dynamic adaptive robust TNEP problem can be written as the following three-level mathematical programming problem. Note that the dual variables are provided after the corresponding constraint separated by a colon.
\begin{equation}\label{eq1}
  \mm{Minimize}{x_k^{(t)}, y_k^{(t)}}{\displaystyle\sum_{t \in \mathcal{T}} \fraca{1}{(1+I)^{t-1}}\left(\displaystyle\sum_{k \in \mathcal{L}^+}c_k y_k^{(t)}+c_{\rm op}^{(t)}\right)}
\end{equation}
subject to
\begin{align}
\Pi & \geq  \displaystyle\sum_{t \in \mathcal{T}} \displaystyle\sum_{k \in \mathcal{L}^+} \fraca{1}{(1+I)^{t-1}}c_k y_k^{(t)}  \label{eq2}\\
  x_k^{(t)} &= 1;\;\forall k \in \mathcal{L}\backslash \mathcal{L}^+,\forall t \in \mathcal{T}\label{eq2a}\\
  x_k^{(t)} &= \sum_{p=1}^{p=t} y_k^{(p)};\;\forall k \in \mathcal{L}^+,\forall t \in \mathcal{T}\label{eq2b}\\
  \sum_{t \in \mathcal{T}} y_k^{(t)} &\leq 1;\;\forall k \in \mathcal{L}^+\label{eq2c}\\
 y_k^{(t)}  & \in  \{0,1\};\;\forall k \in \mathcal{L}^+,\forall t \in \mathcal{T} \label{eq3},
 \end{align}
where operational costs $c_{\rm op}^{(t)}$ in (\ref{eq1}) for each period $t;\;\forall t \in \mathcal{T}$ and for given values of the first-stage decision variables $x_k^{(t)}, y_k^{(t)}$ are obtained by solving the following inner optimization problem:
\begin{align}
c_{\rm op}^{(t)}=\!\!\!\!\mm{Maximum}{{\bfg d}^{(t)}\in \mathcal{{\bfg D}}^{(t)}}\!\!\!\!\!\!\!\!&\mm{Minimum}{g^{(t)}_i, p^{(t)}_j, r^{(t)}_j, \theta^{(t)}_n, f^{(t)}_k}\!\!\!\!\!\!\left(\sigma \displaystyle\sum_{i\in \mathcal{G}^{(t)}}c^{\rm G}_{i}g^{(t)}_{i}+\right. \nonumber \\
&\quad\quad\quad\quad\quad\left.+\sigma \displaystyle\sum_{j\in \mathcal{D}^{(t)}}c^{\rm U}_{j}r^{(t)}_{j}\right),\label{OPC}
\end{align}
subject to
\begin{align}
 \sum_{i \in \Psi_{n}^{\rm G}}g^{(t)}_{i}&-\sum_{k \mid o(k)=n}f^{(t)}_{k}+\sum_{k \mid r(k)=n}f^{(t)}_{k} +\sum_{j \in \Psi_{n}^{\rm D}}r^{(t)}_{j} \nonumber\\
 & = \sum_{j\in \Psi_{n}^{\rm D}}p^{(t)}_{j}: \lambda_{n}^{(t)};\; \forall n \in \mathcal{N}\label{balance}\\
f^{(t)}_{k}&=b_{k}x^{(t)}_{k}(\theta_{o(k)}^{(t)}-\theta_{r(k)}^{(t)}): \phi^{(t)}_{k};\; \forall k \in \mathcal{L} \label{flow}\\
\theta^{(t)}_{n}&=0: \chi^{(t)}_{n};\; n:\text{slack} \label{refer angle}\\
f^{(t)}_{k} &\leq f_{k}^{\rm max}: \hat{\phi}^{(t)}_{k};\; \forall k \in \mathcal{L}\label{flow upper limit}\\
f^{(t)}_{k}&\geq -f_{k}^{\rm max}: \check{\phi}^{(t)}_{k};\; \forall k \in \mathcal{L}\label{flow lower limit}\\
\theta_{n}^{(t)} & \leq \pi: \hat{\xi}^{(t)}_{n};\; \forall n \in \mathcal{N}\backslash n:\text{slack} \label{angle upper limit}\\
\theta_{n}^{(t)} & \geq -\pi: \check{\xi}^{(t)}_{n};\; \forall n \in \mathcal{N}\backslash n:\text{slack} \label{angle lower limit}\\
g^{(t)}_{i} &\geq 0;\; \forall i\in\mathcal{G}^{(t)}  \label{gener lower limit}\\
r^{(t)}_{j}& \geq 0;\; \forall j \in \mathcal{D}^{(t)}\label{loadshed lower limit}\\
p^{(t)}_{j} &= d_{j}^{{\rm D}(t)}: \alpha_{j}^{{\rm D}(t)};\; \forall j \in \mathcal{D}^{(t)}\label{demand upper limit}\\
g^{(t)}_{i}& \leq d_{i}^{{\rm G}(t)}: \varphi_{i}^{{\rm G}(t)};\; \forall i\in \mathcal{G}^{(t)}  \label{gener upper limit}\\
r^{(t)}_{j} &\leq e^{(t)}_j d_{j}^{{\rm D}(t)}: \varphi_{j}^{{\rm D}(t)};\; \forall j\in \mathcal{D}^{(t)}. \label{loadshed upper limit}
\end{align}

The objective function (\ref{eq1}) represents the net present cost (NPC) associated with expansion investment and operational costs, defined as the sum of the present values of costs over the time horizon. Constraint (\ref{eq2}) keeps the maximum amount of expansion investment throughout the time horizon to within the available budget. Constraints (\ref{eq2a}) and (\ref{eq2b}) make the line status  equal to 1 for all existing transmission lines at the beginning of the time horizon considered, and once the line has been constructed, respectively, while constraint (\ref{eq2c}) ensures that no line is constructed more than once throughout the time horizon considered. Constraint (\ref{eq3}) establishes the binary nature of investment decisions. Note that variables ${\bfg x}^{(t)};\;t \in \mathcal{T}$ are also binary, but, these integrality constraints can be relaxed and  variables such as these can be defined as continuous because their binary nature is ensured by means of the set of equations (\ref{eq2a})-(\ref{eq3}).
Equation (\ref{OPC}) represents the least desirable scenario for operational costs made up of maximum generation and load-shedding costs.
Constraint (\ref{balance}) sets the power balance at every bus. Constraint (\ref{flow}) shows the power flow through each line. Note that the power flow depends on the actual status of the line $x^{(t)}_{k}$, thus, if the corresponding line is not physically connected to the network, the power flow through it is zero. Constraint (\ref{refer angle}) fixes the voltage angle of the reference bus to zero.
Constraints (\ref{flow upper limit})-(\ref{flow lower limit}) set the upper and lower line flow limits. Constraints (\ref{angle upper limit})-(\ref{angle lower limit}) set limits on the voltage angles at every bus, and (\ref{gener lower limit})-(\ref{loadshed lower limit}) ensure  the power generation and load-shedding variables are both positive.
Finally, constraint (\ref{demand upper limit}) makes the level of demand match the uncertain demand variable, (\ref{gener upper limit}) sets the power generation to be lower than the uncertain generation capacity variable, and (\ref{loadshed upper limit}) limits load-shedding to a percentage of the uncertain demand variable. The uncertain demand and generation capacity variables are defined in Section \ref{s2}. For purposes of clarity, the number of different uncertainty sets is initially considered equal to the number of years of the time horizon, although different time periods could be considered instead.

It must be stressed that the main difference with respect to previous ARO formulations for TNEP problems is the consideration of different costs for each year throughout the time horizon, which made us include the additional constraints (\ref{eq2a})-(\ref{eq2c}). This is because there are different uncertainty sets for each year ${\bfg D}^{(t)}$, which represent possible changes in loads, generation capacities, etc. The advantage this formulation has over traditional static approaches is that investment decisions can be made at any time throughout the study horizon, providing an integrated representation of the problem. On the downside, computational complexity increases. This is the prize we have to pay to circumvent the simplifying assumptions typically adopted in the static models available in the literature. Nevertheless, it is still a computationally tractable formulation.

\section{Uncertainty Modelling}\label{s2}

Uncertainties that are pertinent to the transmission expansion planning problem in a market setting include:
\begin{enumerate}
  \item demand growth,
  \item spatial distribution of demand growth,
  \item generation capacities
  \item availability of transmission facilities and
  \item availability of generation facilities.
\end{enumerate}

In this paper we only consider uncertainties associated with demand and generation capacities. To be specific, we use the same definition of an uncertainty set as that given by \cite{MinguezG:15} as a starting point, i.e.:
 \begin{eqnarray}
    d^{\rm G}_{i} & = & \bar d^{\rm G}_{i}-\hat d^{\rm G}_{i}z^{\rm G}_{i};\;\forall i\in \mathcal{G} \label{uncer1}\\
    d^{\rm D}_{j} & = & \bar d^{\rm D}_{j}+\hat d^{\rm D}_{j}z^{\rm D}_{j};\;\forall j\in \mathcal{D} \label{uncer3}\\
     \sum_{i\in \mathcal{G}} z^{\rm G}_{i}  &\le &\Gamma^{\rm G}  \\
    \sum_{j\in \mathcal{D}} z^{\rm D}_{j}  &\le &\Gamma^{\rm D}\\
    z^{\rm G}_{i}&\in &\{0,1\};\forall i\in \mathcal{G}\\
    z^{\rm D}_{j}&\in &\{0,1\};\forall j\in \mathcal{D},\label{uncer2}
  \end{eqnarray}
where $d^{\rm G}_{i}$ is the uncertain generation limit for the generating unit $i$, and is related to the $i$th variable within vector ${\bfg d}$,
 $\bar d^{\rm G}_{i}$ is the corresponding nominal value, $\hat d^{\rm G}_{i}$ is the maximum positive distance from the nominal value that can take the random parameter, $z^{\rm G}_{i}$ is an auxiliary variable, and $\Gamma^{\rm G}$ is the maximum number of random parameters for generation capacity which may reach their limits. Likewise, $d^{\rm D}_{j}$, $\bar d^{\rm D}_{j}$, $\hat d^{\rm D}_{j}$, $z^{\rm D}_{j}$ and $\Gamma^{\rm D}$ correspond to the same values but for demand. Note that according to \cite{RuizC:15}, in the least desirable outcome arising from  ``nature'' with  a fixed network configuration, would be one in which there is maximum load shedding and, consequentially, maximum operational costs. Therefore, generation capacity would be as low as possible with respect to nominal values and the demand load as high as possible with respect to nominal values. This explains the signs in (\ref{uncer1})-(\ref{uncer3}).

Since the aim of this paper is to use long-term stochastic processes for TNEP, that might evolve during time horizon, the uncertainty set (\ref{uncer1})-(\ref{uncer2}) must be associated with each year in the study horizon and allow for changes in successive years. Thus, (\ref{uncer1})-(\ref{uncer2}) transforms into the set ${\bfg D}^{(t)}$ for each year $t$ as follows:
%
 \begin{eqnarray}
    d^{{\rm G}{(t)}}_{i}\!\! \!\!\!\!& = &\!\!\!\! \bar d^{{\rm G}(t)}_{i}-\hat d^{{\rm G}(t)}_{i}z^{{\rm G}(t)}_{i};\;\!\!\!\forall i\in \mathcal{G}^{(t)};\!\forall t \in \mathcal{T} \label{uncer1NS}\\
    d^{{\rm D}{(t)}}_{j}\!\!\!\! \!\!& = &\!\!\!\! \bar d^{{\rm D}(t)}_{j}+\hat d^{{\rm D}(t)}_{j}z^{{\rm D}(t)}_{j};\;\!\!\!\forall j\in \mathcal{D}^{(t)} ;\!\forall t \in \mathcal{T}\label{uncer3NS}\\
     \sum_{i\in \mathcal{G}^{(t)}} z^{{\rm G}(t)}_{i}  &\le &\Gamma^{\rm G};\;\forall t   \in \mathcal{T}\\
    \sum_{j\in \mathcal{D}^{(t)}} z^{{\rm D}(t)}_{j}  &\le &\Gamma^{\rm D};\;\forall t \in \mathcal{T}\\
    z^{{\rm G}(t)}_{i}&\in &\{0,1\};\forall i\in \mathcal{G}^{(t)};\forall t \in \mathcal{T}\\
    z^{{\rm D}(t)}_{j}&\in &\{0,1\};\forall j\in \mathcal{D}^{(t)};\forall t \in \mathcal{T}.\label{uncer2NS}
  \end{eqnarray}

In these uncertainty sets, the nominal values $\bar d^{{\rm G}(t)}_{i},\bar d^{{\rm D}(t)}_{j}$ and maximum positive distances from the nominal values $\hat d^{{\rm G}(t)}_{i},\hat d^{{\rm D}(t)}_{j}$ for each year $t$ must be defined. Note that these values represent the nominal or expected value and the dispersion, respectively. Thus, this methodology is highly flexible as it takes different circumstances into account, such as increases and/or decreases in nominal values, increases and/or decreases in dispersion or both.

In addition, both set of indices of demand $\mathcal{D}^{(t)}$ and generation units $\mathcal{G}^{(t)}$ are allowed to change for each year. Thus, it is possible to accommodate expected future consumption nodes, and/or the possible construction of future generation facilities for each year $t$.

Note that all parameters defining uncertainty sets for each year must be defined by planners using expert criterion, or alternatively, the forecasting tools put forward by \cite{NogalesC:06,AlonsoGRS:11,RuizC:15}, such as ARIMA, GARCH, dynamic factors or transfer function models. These tools enable the behaviour of  different demand levels and production capacities to be forecast, after which this data can be used to derive appropriate upper and lower bounds for the uncertainty sets. The difference with respect to the static approach is that those limits are defined for each year considered within the study horizon.

\begin{figure}[htb]
  \begin{center}
  \includegraphics[width=0.5\textwidth]{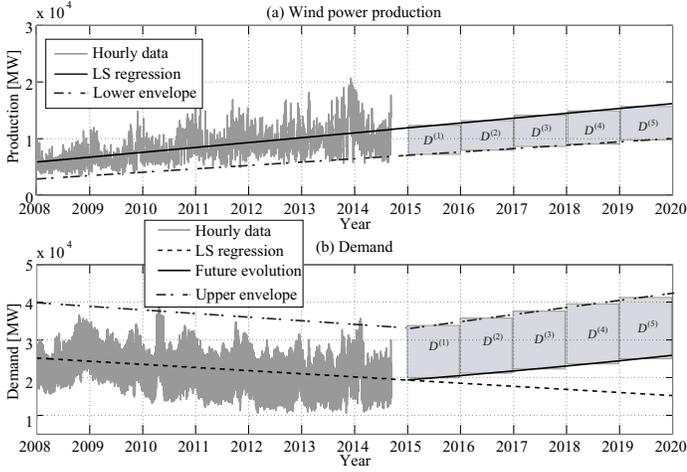}
    \caption{Graphical illustration of the evolution of uncertainty sets associated with Spanish electricity market (OMIE) historical data (2008–-2013).}
    \label{WindDemand}
  \end{center}
\end{figure}

An example of data evolution corresponds to the same data sets considered by \cite{RuizC:15} and is taken from the Spanish electricity market (OMIE, http://www.omie.es/), where wind power production and demand is considered. A least squares (LS) linear regression has been plotted (continuous and dashed black lines respectively) to check possible long term trends associated with nominal or expected values. The change in the installed capacity of wind power results in a positive trend, which has been prolonged until 2020 as shown in Figure~\ref{WindDemand} (a). Moreover, the linear lower production envelope has been traced and prolonged up to 2020. Note that only values below nominal values are considered. Our proposal consists of defining constant uncertainty sets for each year, which are represented by light gray boxes ${\bfg D}^{(1)},\ldots, {\bfg D}^{(5)}$,  that change according to  foreseen nominal and dispersion values of wind power production. Regarding demand, the least squares trend has decreased as a result of the European economic downturn. However, we have used an annual increase rate of 6\%, similar to the rate before the downturn. For demand, only values above nominal values are considered. Note that the aim of this figure is to illustrate the concept of uncertainty set evolution, not to state what the appropriate statistical technique is to make future prognosis (which is beyond the scope of this paper).

Finally, it must be stressed that as with the treatment of uncertain variables within each uncertainty set for static approaches, where the temporal autocorrelation among variables is disregarded because we are only interested in the situation within the uncertainty set in which the highest operational costs occur, no correlation among uncertainty sets associated with each time period (year) is required. In the dynamic approach, our aim is to find the highest operation cost for each uncertainty set independently.

\section{Proposed Decomposition Method}\label{s3}
%

The decomposition method put forward has a bi-level structure, so, the first step is to merge the initial three-level formulation (\ref{eq1})-(\ref{loadshed upper limit}) into a two-level problem. For given values for the first-stage variables ${x}_k^{(t)}$ and ${y}_{k}^{(t)}$ for each line at any year, the problem set out by (\ref{OPC}), (\ref{balance})-(\ref{loadshed upper limit}) might be decomposed by the time period. Thus, the operational cost definition transforms into the following single-level maximization problem for each time period $t;\;\forall t\in \mathcal{T}$:
\begin{align}
&\mm{Maximize}{{\bfg d},\lambda_{n}^{(t)},\phi^{(t)}_{k},\chi^{(t)}_{n},\hat{\phi}^{(t)}_{k},\check{\phi}^{(t)}_{k},\hat{\xi}^{(t)}_{n},\check{\xi}^{(t)}_{n},\alpha_{j}^{{\rm D}(t)},\varphi_{i}^{{\rm G}(t)},\varphi_{j}^{{\rm D}(t)}}{}\nonumber\\
&{\left\{\!\!\!\!\!\!
\begin{array}{c}
 \displaystyle\!\!\!\sum_{k \in \mathcal{L}}\Bigl(\hat{\phi}^{(t)}_{k}-\check{\phi}^{(t)}_{k}\Bigr)f_{k}^{\rm max}\!\!\!+\!\!\!\displaystyle\sum_{n\in \mathcal{N}\backslash n:\textrm{slack}}\pi\Bigl(\hat{\xi}^{(t)}_{n}-\check{\xi}^{(t)}_{n}\Bigr) \\
 +\!\!\!\displaystyle\sum_{i\in \mathcal{G}^{(t)}}\!\!\Bigl(d_{i}^{{\rm G}(t)}\varphi_{i}^{{\rm G}(t)}\Bigr)
 +\!\!\!\displaystyle\sum_{j\in \mathcal{D}^{(t)}}\!\!\Bigl(d_{j}^{{\rm D}(t)}\alpha_{j}^{{\rm D}(t)}\!\!+\!\!e^{(t)}_j d_{j}^{{\rm D}(t)}\varphi_{j}^{{\rm D}(t)}\!\! \Bigr)
\end{array}\!\!\!\!\!\right\}
}
\label{subproblem1}
\end{align}
subject to:
\begin{align}
&\lambda_{n(i)}^{(t)}+\varphi_{i}^{{\rm G}(t)} \leq  \sigma c^{\rm G}_{i}; \;\forall i\in \mathcal{G}^{(t)} \label{dual blockgen}\\
&-\lambda_{n(j)}^{(t)}+\alpha_{j}^{{\rm D}(t)}  \leq  0; \; \forall j \in \mathcal{D}^{(t)} \label{dual blockdemand}\\
&\lambda_{n(j)}^{(t)}+\varphi_{j}^{{\rm D}(t)}  \leq  \sigma c^{\rm U}_{j}; \;\forall j\in \mathcal{D}^{(t)} \label{dual loadshed}\\
&-\lambda_{o(k)}^{(t)}+\lambda_{r(k)}^{(t)}+\phi^{(t)}_{k}+\hat{\phi}^{(t)}_{k}+\check{\phi}^{(t)}_{k} =  0; \; \forall k\in \mathcal{L} \label{dual flow}
\end{align}
\begin{align}
&-\sum_{k\mid o(k)=n}b_{k}x^{(t)}_{k}\phi^{(t)}_{k}+\sum_{k\mid r(k)=n}b_{k}x^{(t)}_{k}\phi^{(t)}_{k}    \nonumber \\
&+\hat{\xi}^{(t)}_{n}+\check{\xi}^{(t)}_{n}  =  0 ; \;\forall n \in \mathcal{N}\backslash n:\text{slack} \label{dual angle}\\
&-\sum_{k\mid o(k)=n}b_{k}x^{(t)}_{k}\phi^{(t)}_{k}+\sum_{k\mid r(k)=n}b_{k}x^{(t)}_{k}\phi^{(t)}_{k}    \nonumber \\
&+\chi^{(t)}_{n} = 0;\;n:\text{slack} \label{dual anglerefer}\\
&-\infty \leq  \lambda_{n}^{(t)}  \leq\infty; \;\forall n \in \mathcal{N} \label{dual lambda}\\
&-\infty \leq \phi^{(t)}_{k}  \leq\infty; \;\forall k \in \mathcal{L} \label{dual phik}\\
&-\infty \leq \chi^{(t)}_{n} \leq\infty; \; n:\text{slack} \label{dual chi}\\
& \hat{\phi}^{(t)}_{k}\leq 0; \;\forall k  \in \mathcal{L}\label{dual phikmax}\\
& \check{\phi}^{(t)}_{k} \geq  0; \;\forall k \in \mathcal{L}\label{dual phikmin}\\
& \hat{\xi}^{(t)}_{n}\leq 0; \;\forall n  \in \mathcal{N}\backslash n:\text{slack}\label{dual ximax}\\
& \check{\xi}^{(t)}_{n} \geq  0; \; \forall n \in \mathcal{N}\backslash n:\text{slack}\label{dual ximin}\\
&-\infty \leq  \alpha_{j}^{{\rm D}(t)}  \leq\infty; \; \forall j \in \mathcal{D}^{(t)} \label{dual alpha}\\
& \varphi_{i}^{{\rm G}(t)} \leq 0; \; \forall i  \in \mathcal{G}^{(t)}\label{dual varphiG}\\
& \varphi_{j}^{{\rm D}(t)} \leq 0; \; \forall i  \in \mathcal{D}^{(t)}\label{dual varphiD}\\
& d^{{\rm G}{(t)}}_{i} = \bar d^{\rm G}_{i}r_{\mu,i}^{(t)}-\hat d^{\rm G}_{i}r_{\sigma,i}^{(t)}z^{{\rm G}(t)}_{i};\;\forall i\in \mathcal{G}^{(t)} \label{uncer1NS_2}\\
& d^{{\rm D}{(t)}}_{j} = \bar d^{\rm D}_{j}r_{\mu,j}^{(t)}+\hat d^{\rm D}_{j}r_{\sigma,j}^{(t)}z^{{\rm D}(t)}_{j};\;\forall j\in \mathcal{D}^{(t)} \label{uncer3NS_2}\\
& \displaystyle\sum_{i\in \mathcal{G}^{(t)}} z^{{\rm G}(t)}_{i}  \le \Gamma^{\rm G}\\
& \displaystyle\sum_{j\in \mathcal{D}^{(t)}} z^{{\rm D}(t)}_{j}  \le \Gamma^{\rm D}\\
& z^{{\rm G}(t)}_{i}\in \{0,1\};\forall i\in \mathcal{G}^{(t)}\\
& z^{{\rm D}(t)}_{j}\in \{0,1\};\forall j\in \mathcal{D}^{(t)}.\label{uncer2NS_2}
\end{align}

Subproblems (\ref{subproblem1})-(\ref{uncer2NS_2}) result from substituting the third-level problem by its dual and incorporation of definition equations for the uncertainty set (\ref{uncer1NS_2})-(\ref{uncer2NS_2}), and there are the same number of subproblems as there are years under consideration $n_y$ within the time horizon. These subproblems provide the uncertain parameter values ${\bfg d}^{(t)}$ within the uncertainty sets to give the least desirable operational costs for each year.

The only additional detail required in order to define the subproblems properly is linealization of the bilinear terms included in (\ref{subproblem1}), i.e. $\sum_{i\in \mathcal{G}^{(t)}}\Bigl(d_{i}^{{\rm G}(t)}\varphi_{i}^{{\rm G}(t)}\Bigr) +\sum_{j\in \mathcal{D}^{(t)}}\Bigl(d_{j}^{{\rm D}(t)}\alpha_{j}^{{\rm D}(t)}+e^{(t)}_j d_{j}^{{\rm D}(t)}\varphi_{j}^{{\rm D}(t)} \Bigr)$. Taking into account equations (\ref{uncer1NS_2})-(\ref{uncer3NS_2}), this bilinear term becomes:
\begin{align}\label{lineali}
&\displaystyle\sum_{i\in \mathcal{G}^{(t)}}\Bigl(d_{i}^{{\rm G}(t)}\varphi_{i}^{{\rm G}(t)}\Bigr)\!\! +\!\!\displaystyle\!\!\sum_{j\in \mathcal{D}^{(t)}}\Bigl(d_{j}^{{\rm D}(t)}\alpha_{j}^{{\rm D}(t)}\!\!+\!\!e^{(t)}_j d_{j}^{{\rm D}(t)}\varphi_{j}^{{\rm D}(t)} \Bigr)\nonumber \\
&=\displaystyle\sum_{i\in \mathcal{G}^{(t)}}\Bigl(\bar d^{\rm G}_{i}r_{\mu,i}^{(t)}\varphi_{i}^{{\rm G}(t)}-\hat d^{\rm G}_{i}r_{\sigma,i}^{(t)}z^{{\rm G}(t)}_{i}\varphi_{i}^{{\rm G}(t)}\Bigr) \nonumber\\
&+\displaystyle\sum_{j\in \mathcal{D}^{(t)}}\Bigl(\bar d^{{\rm D}(t)}_{j}r_{\mu,j}^{(t)}\alpha_{j}^{{\rm D}(t)}+\hat d^{{\rm D}(t)}_{j}r_{\sigma,j}^{(t)}z^{{\rm D}(t)}_{j}\alpha_{j}^{{\rm D}(t)}\Bigr)\nonumber\\
&+\displaystyle\sum_{j\in \mathcal{D}^{(t)}}e^{(t)}_j\Bigl(\bar d^{{\rm D}(t)}_{j}r_{\mu,j}^{(t)}\alpha_{j}^{{\rm D}(t)}+\hat d^{{\rm D}(t)}_{j}r_{\sigma,j}^{(t)}z^{{\rm D}(t)}_{j}\alpha_{j}^{{\rm D}(t)}\Bigr).
\end{align}

The terms to be linearized correspond to products of binary and dual variables, $z^{{\rm G}(t)}_{i}\varphi_{i}^{{\rm G}(t)}$ and $z^{{\rm D}(t)}_{j}\alpha_{j}^{{\rm D}(t)}$. Details about the technique used for  linearization  are divulged in \cite{MinguezG:15}. The resulting formulation associated with subproblems is a mixed-integer linear programming problem, which can be solved efficiently by using state-of-the-art mixed-integer mathematical programming solvers such as CPLEX or Gurobi.

The optimal solutions for subproblems (\ref{subproblem1})-(\ref{uncer2NS_2}) provide the uncertain parameter values ${\bfg d}^{(t)}$ for each year in order to construct primal cuts for the master problem, which corresponds to the following optimization problem at iteration $\nu$:
\begin{equation}\label{master5}
  \mm{Minimize\!\!\!\!\!\!\!\!\!\!\!\!}{\!\!\!\!\!\!\!\!\!\!\!\begin{array}{c}
                  x_k^{(t)}, y_k^{(t)}, g^{(t)}_{i,l}, p^{(t)}_{j,l}, \\
                  r^{(t)}_{j,l}, \theta^{(t)}_{n,l}, f^{(t)}_{k,l} \\
                  \forall t \in \mathcal{T}\\ l= 1,\ldots,\nu-1
                \end{array}}{\displaystyle\sum_{t \in \mathcal{T}} \fraca{1}{(1+I)^{t-1}}\left(\displaystyle\sum_{k \in \mathcal{L}^+}c_k y_k^{(t)}+\gamma^{(t)}\right)}
\end{equation}
subject to
\begin{align}
\gamma^{(t)} & \ge  \sigma \displaystyle\sum_{i\in \mathcal{G}^{(t)}}c^{\rm G}_{i}g^{(t)}_{i,l}+ \sigma \displaystyle\sum_{j\in \mathcal{D}^{(t)}}c^{\rm U}_{j}r^{(t)}_{j,l};\;\nonumber\\
&\quad \quad\quad\quad\forall t \in \mathcal{T}, l = 1,\ldots,\nu-1\label{master7}\\
\gamma^{(t)}& \ge 0;\;\forall t\in \mathcal{T}\\
\Pi & \geq  \displaystyle\sum_{t \in \mathcal{T}} \displaystyle\sum_{k \in \mathcal{L}^+} \fraca{1}{(1+I)^{t-1}}c_k y_k^{(t)}  \label{eq2_m}\\
  x_k^{(t)} &= 1;\;\forall k \in \mathcal{L}\backslash \mathcal{L}^+,\forall t \in \mathcal{T}\label{eq2a_m}\\
  x_k^{(t)} &= \sum_{p=1}^{p=t} y_k^{(p)};\;\forall k \in \mathcal{L}^+,\forall t \in \mathcal{T}\label{eq2b_m}\\
  \sum_{t \in \mathcal{T}} y_k^{(t)} &\leq 1;\;\forall k \in \mathcal{L}^+\label{eq2c_m}\\
 y_k^{(t)}  & \in  \{0,1\};\;\forall k \in \mathcal{L},\forall t \in \mathcal{T} \label{eq3_m}\\
 \textrm{Equations } &(\ref{balance})-(\ref{loadshed upper limit});\;l = 1,\ldots,\nu-1.\label{masterfinal}
\end{align}

Note that the master problem, besides variables $\gamma^{(t)}$ relates to year on year operational costs, includes one variable $g^{(t)}_{i,l}$, $p^{(t)}_{j,l}$, $r^{(t)}_{j,l}$, $\theta^{(t)}_{n,l}$ and $f^{(t)}_{k,l}$ for each year and for each realization of the uncertain parameters obtained from the subproblem (\ref{subproblem1})-(\ref{uncer2NS_2}) at every iteration. As pointed out by \cite{RuizC:15} in the static approach, the master problem does not pose any computational challenge since it only incorporates a small number of primal cuts (small number of iterations $\nu$). However, in the dynamic method, the number of primal cuts and first-level binary variables is multiplied by the number of time periods (years) under consideration, thereby increasing exponentially the computational time required to solve the master problem. Alternatively, instead of working with time periods in years, longer periods could be used. The optimal time period  must be long enough to reduce the number of subproblems and, therefore, the computational complexity and short enough for the uncertainty set to accurately display the non-stationary nature of the uncertain variables. In other words, there must be a suitable tradeoff between complexity and display of the non-stationary characteristics. This fundamentally depends on how fast those parameters might change throughout the time horizon. Nevertheless, appropriate selection of this period is beyond the scope of this paper.

Once the subproblems and master problem formulations are given, the solution method consists of iteratively solving the following problems:
\begin{itemize}
  \item {\bf Subproblems, one for each year:} For given values for the first-stage variables $x_k^{(t)}$ and $y_k^{(t)}$, the subproblems in  (\ref{subproblem1})-(\ref{uncer2NS_2}) obtain the values for the uncertain parameters within the uncertainty set to obtain the least desirable operational costs (\ref{OPC}).
  \item {\bf Master problem:} Given the least desirable realizations of the uncertain parameters in terms of operational costs, new values for the first-stage variables $x_k^{(t)}$ and $y_k^{(t)}$ are calculated by means of (\ref{master5})-(\ref{masterfinal}).
\end{itemize}

The proposed iterative scheme put forward is described step by step in the following algorithm:
\begin{algorithm}{Dynamic robust transmission network expansion planning}
\begin{description}
\item[{\bf Input:}]\hspace*{0.2cm}Selection of uncertainty budgets $\Gamma^{\rm G}$ and $\Gamma^{\rm D}$, time periods to divide the time horizon, interest rate $I$, definition of the uncertainty sets for each time period, and tolerance of the process $\varepsilon$. These data are selected by the decision maker.

\item[{\bf Step 1:}]\hspace*{0.2cm}{\bf Initialization.} Initialize the iteration counter to $\nu=1$, and upper and lower bounds of the objective function $z^{(\rm up)} = \infty$ and $z^{(\rm lo)} = -\infty$.

\item[{\bf Step 2:}]\hspace*{0.2cm}{\bf Solving the master problem at iteration $\nu$.} Solve the master problem (\ref{master5})-(\ref{masterfinal}). The result provides values of the decision variables $x_{k,\nu}^{(t)}$, $y_{k,\nu}^{(t)}$ and $\gamma^{(t)}$. Update the optimal objective function lower bound $z^{(\rm lo)}=\sum_{t\in\mathcal{T}} \frac{1}{(1+I)^{t-1}}\left(\sum_{k \in \mathcal{L}}c_k y_{k,\nu}^{(t)}+\gamma^{(t)}\right)$. Note that at the first iteration the optimal solution matches the no investment case. Alternatively, we could start with any other vector for decision variables.

\item[{\bf Step 3:}]\hspace*{0.2cm}{\bf Solving subproblems at iteration $\nu$ for each year $t$.} For given values of the decision variables $x_{k,\nu}^{(t)}$, $y_{k,\nu}^{(t)}$, we calculate the least desirable operational costs within the uncertainty set $c_{{\rm op},\nu}^{(t)}$, whereby we also obtain the corresponding uncertain parameters ${\bfg d}^{(t)}_{\nu}$. This is achieved by solving subproblems (\ref{subproblem1})-(\ref{uncer2NS_2}).
    Update the optimal objective function upper bound $z^{(\rm up)}=\sum_{t\in \mathcal{T}} \frac{1}{(1+I)^{t-1}}\left(\sum_{k \in \mathcal{L}}c_k y_{k,\nu}^{(t)}+c_{{\rm op},\nu}^{(t)}\right)$.

\item[{\bf Step 4:}]\hspace*{0.2cm}{\bf Convergence checking.} If $(z^{(\rm up)}-z^{(\rm lo)})/z^{(\rm up)}\le \varepsilon$ go to {\em Step 5}, else update the iteration counter $\nu\rightarrow \nu+1$ and continue from {\em Step 2}.

\item[{\bf Step 5:}]\hspace*{0.2cm}{\bf  Output.} The solution for a given tolerance corresponds to $y_{k}^{(t)\ast}=y_{k,\nu}^{(t)}$.

\end{description}
\end{algorithm}

It must be stressed that the resulting bi-level formulation given by (\ref{subproblem1})-(\ref{uncer2NS_2})  and (\ref{master5})-(\ref{masterfinal}) has the same problem structure than that defined by \cite{ZengZ:13}, and therefore the column-and-constraint generation method used to solve the problem ensures convergence to a global optimum.

\section{Examples}\label{CaseStudy}
It is worth stressing that we have selected two types of analysis to be presented in this section:
\begin{enumerate}
  \item For the first analysis, examples have been strategically designed to enable both static and dynamic approaches to be comparable.  We have assumed that the network must withstand ever higher traffic loads which always increase by the same proportion. Therefore, the least desirable situation from a capacity design perspective, i.e. one in which there are the highest operational costs, is that which corresponds to the last year $n_y$ in the study horizon.
The uncertainty set associated with this last year is the one required to carry out a static analysis. Unlike static analysis, with the dynamic approach it is assumed that the same final design situation will be reached at the end of the study horizon but the variation of uncertain parameters is distributed throughout the study period. This is important because since expansion planning is conditioned by the uncertainty set at the end of the study horizon, i.e. assuming that is the one in which the highest operational costs among all the uncertainty sets occurs, then the solution in terms of expanding lines at the end of the study horizon for both approaches must be the same.
  \item The second analysis attempts to solve the same dynamic problem from the first analysis but using a sequential static approach.
  Thus, the selected target years are treated in a separate and sequential way starting from year one, i.e. TNEP problems associated with different years are solved sequentially assuming that the lines chosen to be constructed in a particular year are available for posterior periods. The aim of this analysis is to contrast the global character associated with the solution given by the proposed method in comparison with the sequential static approach.
\end{enumerate}

\subsection{Illustrative Case Study. Garver System}
\begin{figure}[htb]
  \begin{center}
  \includegraphics[width=0.5\textwidth]{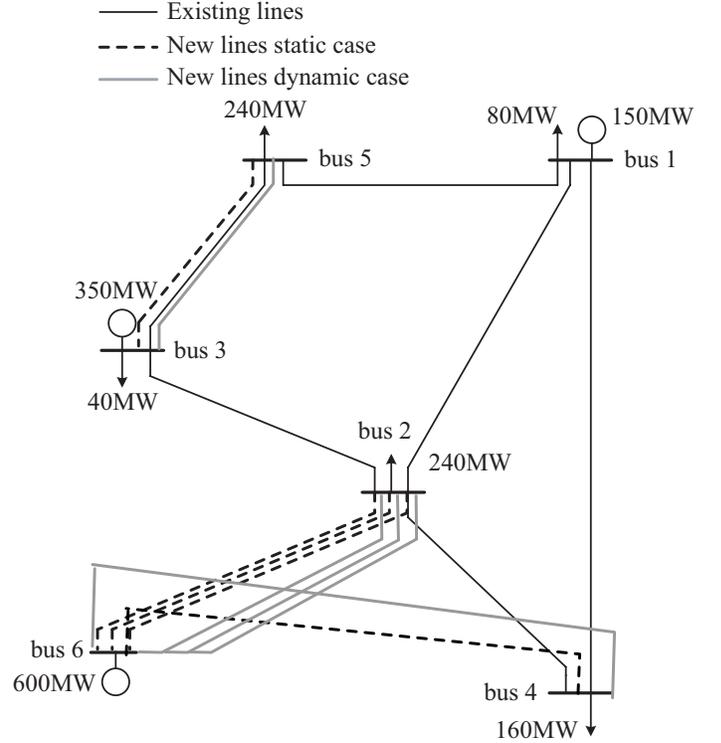}
    \caption{Garver's 6-bus test system.}
    \label{Figure1}
  \end{center}
\end{figure}

The model put forward is illustrated with the Garver 6-bus system, depicted in Figure~\ref{Figure1}. This system is made up of 6 buses, 3 generators, 5 levels of inelastic demand and 6 lines. Nominal values for generation capacities and demand and their supply and bidding prices can be found in \cite{MinguezG:15}. The load-shedding cost is equal to the bidding price for each level of demand.
It has been thought that up to three lines could be installed between each pair of buses. Line data are obtained from Table I of \cite{GarcesCGR:09}, including construction costs, and the maximum available investment budget is \ 40 million euros.

The time horizon is thought to be 25 years, and the discount rate is 10\%. Since the pseudo-stationary periods are deemed to be one year, the weighted factor $\sigma$ associated with load-shedding and power generation costs is equal to the number of hours in a year, i.e. 8760.

In order to compare results yielded from this method with respect to the static approach, we can consider the same example given by \cite{MinguezG:15}, where power generation capacities can increase or decrease by up to 50\% of their nominal values, while demand levels may change by up to  20\%. We assume that the uncertainty set associated with the first year, including nominal values and deviations from nominal values, is the same as that given in \cite{MinguezG:15} but reduced by 25\%. For the remaining years, we assume that the annual growth rates for nominal values and deviations are equal to 1.2\% so that the uncertainty set parameters are equal to:
\begin{equation}\label{GarverUSVar}
\begin{array}{rcl}
 \bar d^{{\rm G}(t)}_{i}\!\!\!&\!\!\!=\!\!\!&\!\!\!(1+0.012)\bar d^{{\rm G}(t-1)}_{i};\;\forall i\in\mathcal{G}^{(t)};\forall t \in \mathcal{T} \wedge t>1\\
 \bar d^{{\rm D}(t)}_{j}\!\!\!&\!\!\!=\!\!\!&\!\!\!(1+0.012)\bar d^{{\rm D}(t-1)}_{j};\;\forall j\in\mathcal{D}^{(t)};\forall t \in \mathcal{T} \wedge t>1\\
 \hat d^{{\rm G}(t)}_{i}\!\!\!&\!\!\!=\!\!\!&\!\!\!(1+0.012)\hat d^{{\rm G}(t-1)}_{i};\;\forall i\in\mathcal{G}^{(t)};\forall t \in \mathcal{T} \wedge t>1\\
 \hat d^{{\rm D}(t)}_{j}\!\!\!&\!\!\!=\!\!\!&\!\!\!(1+0.012)\hat d^{{\rm D}(t-1)}_{j};\;\forall j\in\mathcal{D}^{(t)};\forall t \in \mathcal{T} \wedge t>1.
\end{array}
\end{equation}

Using these rates of change the uncertainty set defined for year 25 is equal to the uncertainty set given in \cite{MinguezG:15}. Note that this last uncertainty set corresponds to the least desirable possible outcome from a transmission network expansion perspective, i.e. the one in which the highest operational costs occur.

Regarding the first type of analysis and using the uncertainty budgets $\Gamma^{G}=2$ and $\Gamma^{D}=2$, in the solution given by the static approach \cite{MinguezG:15} a total of \ 27.031 million euros is invested for constructing the lines shown in Figure~\ref{Figure1} (dashed black lines), these lines are constructed at the beginning of the study period. Conversely, in the dynamic solution \ 22.775 million euros is invested for constructing the lines shown in Figure~\ref{Figure1} (gray lines). Both constructed lines are exactly the same in terms of  their optimal solutions, although with the dynamic approach the rate of investment is lower because three of the lines have been constructed at the beginning of the time horizon, while one of the lines joining nodes 2 and 6 is constructed for year 6 and the line between nodes 3 and 5 is constructed for year 9. In terms of the objective function (\ref{eq1}), optimal values for both approaches correspond to \ 188005.544 and \ 188001.288 million euros, respectively, which means that the static solution is more expensive with respect to the dynamic solution.

Note that at the end of the study period, the final solution in terms of new lines given by the static and dynamic approaches is the same as expected, as it is dominated by the last uncertainty set, which is the one in which the highest operational costs occur and the one used for the static approach.
%
This is a particular case strategically selected by the authors in order to compare both approaches in a more meaningful way.
To date the advantage the dynamic approach has is associated with the cost of money over time, which allows  the overall costs to be reduced. However, this is not the only advantage. Let us assume that in year 8, an analysis of the data yielded throughout those seven years reveals that the 1.2\% trend which was initially assumed would be needed for design was higher than in reality, then it would be possible to re-run the model incorporating this new information so that corrective actions could be taken. Let us consider that growth rates throughout those 7 years were indeed $0.4$\% instead of 1.2\%. If we re-ran the model for the remaining years using said 0.4\%  rate of growth, the results would confirm that the line between nodes 3 and 5 would not be necessary, by which we could save an additional \ 1.8015 million euros. This is possible for the dynamic solution since those lines have not been constructed yet. However, even though the static transmission network expansion planning is oversized, no correction actions are possible because those lines were already constructed. The dynamic alternative enables more rational use of existing financial resources making the model less prone to wrongful design assumptions.

Note that these situations are not unusual, as shown in Figure\ref{WindDemand} (b). Figure\ref{WindDemand} (b) depicts aggregate demand in Spain from 2008 to 2015. It can be seen that from 2009 to 2015 the growth trend stops and demand starts decreasing due to the economic downturn. Note that a static design using data from 2008 would have probably resulted in oversized expansion planning. However, the dynamic approach enables corrective measures to be carried out in order to adapt to the real change in uncertain variables.

For the second type of analysis, we solve the sequential static approach using the same data as that for the dynamic problem. In this particular case, the cost of construction is \ 22.611 million euros, i.e. cheaper than the dynamic solution, however, in terms of the objective function (\ref{eq1}), the sequential optimal solution value is \ 188002.146 million euros, which is more expensive than the dynamic solution of \ 188001.288 million euros. In terms of new lines both solutions are equal, the only difference occurs in the line between nodes 3 and 5, which is constructed for year 10 in the sequential solution and for year 9 in the dynamic solution. Anticipating one year the construction of line between nodes 3 and 5 increases construction but reduces operational costs so that the final total cost is reduced.
 This confirms that the sequential approach does not provide a global solution, which is given by the dynamic approach.

Additional observations regarding computational tractability are pertinent:
\begin{enumerate}
 \item Computational times for the static, sequential static and dynamic versions are, respectively, $0.842$, $8.5$ and $68.701$ seconds respectively using a Microsoft Windows Server 2012 with four processors clocking at 2.00GHz. Note that the dynamic assumption increases computational times considerably with respect to the static and sequential approaches, however, it ensures the achievement of the global optimal solution.

  \item Both the static and dynamic approaches require the same number of iterations, i.e. four. The sequential approach requires three iterations for 24 out of 25 problems related to each year, while the remainder problem requires four iterations.

  \item The master problem computational complexity associated with each approach is provided in Table~\ref{payoff}. Note that the possibility of making investments at any time within the project horizon increases the discrete variable number by the number of time periods considered, i.e. $45\times 25=1125$. This explains the increased complexity.
      \end{enumerate}

\begin{table}[ht]
    \caption{Computational complexity of the master problem related to Garver's 6-bus test system illustrative example.}
    \label{payoff}
    \centering
     \renewcommand{\tabcolsep}{1mm}
\begin{tabular}{cccc}
\hline
&  $\sharp$ cont. variables & $\sharp$ discrete variables & $\sharp$ equations \\
\hline
Static & $120$ & $45$ & $220$ \\\hline
Sequential& $238$ & $45$ & $516$ \\\hline
Dynamic & $7751$ & $1125$  & $17216$\\\hline
\end{tabular}
\end{table}

\subsection{Case study. IEEE 118-bus test system}
We run additional computational tests using the more realistic IEEE 118-bus test system~\cite{PSTCA:13}.
The system is made up of 118 buses, 186 existing lines, $54$ generating units and $91$ loads. Additionally, and as with the example given in \cite{MinguezG:15}, it is possible to construct up to 61 additional lines to duplicate each one of the following existing lines: 8, 12, 23, 32, 38, 41, 51, 68, 78, 96, 104, 118, 119, 121, 125, 129, 134, 159, 7, 9, 36, 117, 71, 131, 133, 147, 103, 65, 144, 168, 4, 13, 132, 69, 66, 67, 5, 89, 29, 167, 145, 70, 42, 90, 16, 174, 98, 99, 185, 93, 94, 128, 164, 97, 153, 146, 116, 163, 31, 92, 130.
Data for lines in existing corridors are taken from \cite{PSTCA:13}.
The investment budget is \ 100 million euros, the time horizon period considered is 10 years, and the discount rate is 10\%.
Data for generation capacities and demand loads are given in \cite{MinguezG:15}. The load-shedding cost equals the bidding price of each level of demand multiplied by 1.2

As with the illustrative example, and in order to compare results from this method with respect to the static method, the same example given by \cite{MinguezG:15} is considered, where power generation capacities and levels of demand can increase or decrease by up to 50\% of their nominal values. We assume for the first year the same uncertainty set considered in \cite{MinguezG:15} but reduce its expected value and interval range by two thirds. For the remaining years, we assume that the growth rates of nominal values and deviations are equal to 3.25\% so that the uncertainty set parameters are equal to:
\begin{equation}\label{GarverUSVar2}
\begin{array}{rcl}
  \bar d^{{\rm G}(t)}_{i}\!\!\!&\!\!\!=\!\!\!&\!\!\!(1+0.0325)\bar d^{{\rm G}(t-1)}_{i};\;\forall i\in\mathcal{G}^{(t)};\forall t \in \mathcal{T} \wedge t>1\\
 \bar d^{{\rm D}(t)}_{j}\!\!\!&\!\!\!=\!\!\!&\!\!\!(1+0.0325)\bar d^{{\rm D}(t-1)}_{j};\;\forall j\in\mathcal{D}^{(t)};\forall t \in \mathcal{T} \wedge t>1\\
 \hat d^{{\rm G}(t)}_{i}\!\!\!&\!\!\!=\!\!\!&\!\!\!(1+0.0325)\hat d^{{\rm G}(t-1)}_{i};\;\forall i\in\mathcal{G}^{(t)};\forall t \in \mathcal{T} \wedge t>1\\
 \hat d^{{\rm D}(t)}_{j}\!\!\!&\!\!\!=\!\!\!&\!\!\!(1+0.0325)\hat d^{{\rm D}(t-1)}_{j};\;\forall j\in\mathcal{D}^{(t)};\forall t \in \mathcal{T} \wedge t>1.
\end{array}
\end{equation}

Use of these ratios make the uncertainty set defined for year 10  equal to the uncertainty set given in \cite{MinguezG:15} but with the nominal values and interval range reduced by one third. Generation and demand nominal values and their possible deviations grow linearly reaching their maximum values at the end of the study period. Note that the uncertainty set for year $n_y=10$ corresponds to the least desirable outcome possible from a transmission network expansion perspective, i.e. the uncertainty set in which the maximum operational costs occur among all uncertainty sets.

Regarding the first type of analysis and using the uncertainty budgets $\Gamma^{\rm G}=15$ and $\Gamma^{\rm D}=20$, in the solution given by the static approach \cite{MinguezG:15} (reducing also the nominal values and interval ranges by one third) a total of \ 79.384 million euros are invested for constructing the lines 7, 8, 9, 38, 41, 133, 134, 153, 159. These lines are constructed at the beginning of the study period. Conversely, with the dynamic solution  \ 73.626 million euros are invested for constructing exactly the same lines as in the static solution. However, with the dynamic approach investment is lower because lines 38 and 133 are planned to be constructed for year 3 and line 41 is constructed for year 8. In terms of the objective function (\ref{eq1}), optimal values for both approaches correspond to \ 87905.017 and \ 87899.258 million euros, respectively, which means that the static solution is more expensive with respect to the dynamic solution.

Note that as in the previous example, the advantage the dynamic approach has so far is related to the cost of money  over time, because both solutions are conditioned by the last uncertainty set. Due to this it can be confirmed that the dynamic model provides consistent solutions. However, let us assume that in year 7, an analysis of data throughout those seven years reveals that the 3.25\% trend initially assumed for design was higher than the actual one, it would then be possible to re-run the model with this new information added so that corrective actions can be taken. Let us consider that growth rates throughout those 7 years were indeed below $2.52$\% instead of 3.25\%. If we re-ran the dynamic model for the remaining years using, for instance, 2.4\% growth rates, the results would confirm line 41 to be unnecessary, with which we could save an additional \  2.0285 million euros. This is possible because those lines have not been constructed yet.

For the second type of analysis, we solve the sequential static approach using the same data as that for the dynamic problem. In this particular case, the cost of construction is \ 70.762 million euros, i.e. cheaper than the dynamic solution, however, in terms of the objective function (\ref{eq1}), the sequential optimal solution value is \ 87907.945 million euros, which is more expensive than the dynamic solution of \ 87899.258 million euros. In terms of new lines both solutions are equal, the only difference occurs for lines 38, 41 ad 133, which are constructed for years five, three and five, respectively, in the sequential solution and for years three, eight and three, respectively, in the dynamic solution. These different timing associated with the dynamic solution increases construction but reduces operational costs so that the final total cost is reduced. This confirms that the sequential approach does not provide a global solution, which is given by the dynamic approach.

Additional observations regarding computational tractability are pertinent:
\begin{enumerate}
  \item Computational times for the static, sequential static and dynamic versions are $18.563$, $68.439$ and $366.245$ seconds respectively using a Microsoft Windows Server 2012 with four processors clocking at 2.00GHz. Note that the dynamic assumption increases computational times considerably with respect to the static and sequential approaches, which is an expected result according to the increment in complexity. Nevertheless, the dynamic approach ensures the achievement of the global optimal solution.

  \item In the static approach six iterations are required, while in the dynamic case eight iterations are needed to reach the optimal solution. The sequential approach requires between three and eight iterations depending on the year.

  \item The master problem computational complexity associated with each model is provided in Table~\ref{payoff2}. Note that the possibility of making investments at any time within the project horizon increases the discrete variable number by the number of time periods considered, i.e. $247\times 10=2470$ which can explain the increase in computational time.

      \begin{table}[ht]
    \caption{Master problem computational complexity related to  IEEE 118-bus test system example. }
    \label{payoff2}
    \centering
     \renewcommand{\tabcolsep}{1mm}
\begin{tabular}{cccc}
\hline
&  $\sharp$ cont. variables & $\sharp$ discrete variables & $\sharp$ equations \\
\hline
Static & $1,743$ & $247$ & $3,346$ \\\hline
Sequential & $5,725$ & $247$ & $11,549$ \\\hline
Dynamic & $57,241$ & $2470$  & $114,923$\\\hline
\end{tabular}
\end{table}

   \item The change in the algorithm during the solution process associated with the static approach is given in Table~\ref{Statio118}. Note how the change in the lower and upper bounds tend to converge to the same value, and how investment costs $c_{{\rm in},\nu}$ change among iterations.

     \begin{table}[ht]
    \caption{Change in the static scenario algorithm for the IEEE 118-bus test system example. }
    \label{Statio118}
    \centering
     \renewcommand{\tabcolsep}{1mm}
\begin{tabular}{ccccccc}
\hline
$\nu$ &  1 & 2 & 3 & 4 & 5 & 6\\
\hline
$c_{{\rm op},\nu}$ &  18718.57  & 15263.49  & 15499.74 & 15369.24 & 15317.92 & 15263.49\\
$c_{{\rm in},\nu}$ &  0         & 97.817    & 19.416   & 57.294   & 68.154   & 79.384\\
$z^{(\rm up)}$           &  18718.57  & 15273.27  & 15273.27 & 15273.27 & 15273.27 & 15271.43 \\
$z^{(\rm lo)}$           &  $-\infty$ & 13807.49  & 15265.43 & 15269.29 & 15270.31 & 15271.43 \\
error$^{(\S)}$                      & 1          & 0.09597   & 0.000513 & 0.000260 & 0.000194 & 0.0 \\\hline
\multicolumn{7}{l}{Units in million \euro}\\
\multicolumn{7}{l}{$(\S)$: Adimensional value}\\
\end{tabular}
\end{table}

       \item The change in the algorithm during the solution process associated with the dynamic scenario is given in Table~\ref{NonStatio118}. The change in lower and upper bounds tend to converge on the same value, due to the variations in investment costs $c_{{\rm in},\nu}$ among iterations. It is also important to verify in the Table that the operational costs for the last year $n_y=10$ at each iteration are the highest, because this is the most critical uncertainty set. Additionally, it should be noted how the operation costs for the last year $n_y=10$ at the beginning of the iterative process (\ 18718.57 million euros) and at the end  (\ 15263.49 million euros) match the operational costs found with the static solution.

      \end{enumerate}

\begin{table}[ht]
    \caption{Evolution of the dynamic case algorithm for the IEEE 118-bus test system example. }
    \label{NonStatio118}
    \centering
     \renewcommand{\tabcolsep}{0.5mm}
     {\scriptsize
\begin{tabular}{cccccccc}
\hline
$\nu$ &  1 & 2 & 3 & $\ldots$ & 6 & 7 & 8\\
\hline
$c_{{\rm op},\nu}^{(1)}$ & 14038.928 & 11405.560 & 11558.403 & $\ldots$ & 11453.623 & 11407.980 & 11405.560\\
$c_{{\rm op},\nu}^{(2)}$ & 14558.888 & 11830.358 & 11948.694 & $\ldots$ & 11830.358 & 11830.358 & 11830.358\\
$c_{{\rm op},\nu}^{(3)}$ & 15078.848 & 12255.157 & 12392.574 & $\ldots$ & 12264.433 & 12257.387 & 12255.157\\
$c_{{\rm op},\nu}^{(4)}$ & 15598.809 & 12679.956 & 12836.455 & $\ldots$ & 12710.111 & 12689.654 & 12679.956\\
$c_{{\rm op},\nu}^{(5)}$ & 16118.769 & 13104.755 & 13284.671 & $\ldots$ & 13104.755 & 13104.755 & 13104.755\\
$c_{{\rm op},\nu}^{(6)}$ & 16638.729 & 13529.554 & 13735.844 & $\ldots$ & 13529.554 & 13529.554 & 13529.554\\
$c_{{\rm op},\nu}^{(7)}$ & 17158.690 & 13961.002 & 14168.095 & $\ldots$ & 13961.002 & 13961.002 & 13961.002\\
$c_{{\rm op},\nu}^{(8)}$ & 17678.650 & 14395.165 & 14611.976 & $\ldots$ & 14395.165 & 14395.165 & 14395.165\\
$c_{{\rm op},\nu}^{(9)}$ & 18198.610 & 14829.328 & 15055.856 & $\ldots$ & 14829.328 & 14829.328 & 14829.328\\
$c_{{\rm op},\nu}^{(10)}$& 18718.570 & 15263.492 & 15504.033 & $\ldots$ & 15263.492 & 15263.492 & 15263.492\\\hline
$c_{{\rm in},\nu}$       & 0.0        & 99.178   & 9.796      & $\ldots$   & 69.359    & 71.081    & 73.625 \\
$z^{(\rm up)}$           & 107982.298 & 87924.811 & 87924.811 & $\ldots$ & 87924.811 & 87908.264 & 87899.258\\
$z^{(\rm lo)}$           & $-\infty$  & 78666.752 & 87835.871 & $\ldots$& 87894.992 & 87896.714 & 87899.258\\
error$^{(\S)}$                    & 1          & 0.1053    & 0.0010    &$\ldots$   & 0.0003    & 0.0001    & 0.0000\\\hline
\multicolumn{8}{l}{Units in million \euro}\\
\multicolumn{8}{l}{$(\S)$: Adimensional value}\\
\end{tabular}}
\end{table}

Finally, it is worth stressing that the sequential static and dynamic approaches are more flexible in so far as they can accommodate different uncertainty sets associated with uncertain parameters for different time periods, however, only the dynamic solution ensures the achievement of the optimal global solution.

\section{Conclusions}\label{Conclusions}
In this paper the use of robust optimization for solving the dynamic transmission expansion planning problem has been extended, which is more realistic as regards the variability of energy resources and requirements. The model put forward herein provides the initial design and the expansion plan as regards forthcoming years, assuming that the probability distributions for the random variables (uncertainty sets) change between consecutive years. The proposed model provides an integrated approach reaching the global optimal solution, which circumvents the simplifying assumptions and/or heuristic solutions typically adopted in the {\it static} and {\it sequential static} models available in the literature.

In summary, the proposed method overcomes the size limitations and computational intractability associated with dynamic TNEP for realistic cases, addressing a yet unresolved and challenging problem which is of utmost practical interest for keeping the full dynamic complexity of the problem. This allows to benefit from the goodness of dynamic models, such as corrective actions can be carried out throughout the study horizon, especially in outcomes in which the future variable prognosis differs from true evolution. This alternative enables more rational use of existing financial resources and the model is less prone to wrongful design assumptions.

\ifCLASSOPTIONcaptionsoff
  \newpage
\fi



\vspace{-1cm}

\begin{IEEEbiography}[{\includegraphics[width=1in,height=1.25in,clip,keepaspectratio]{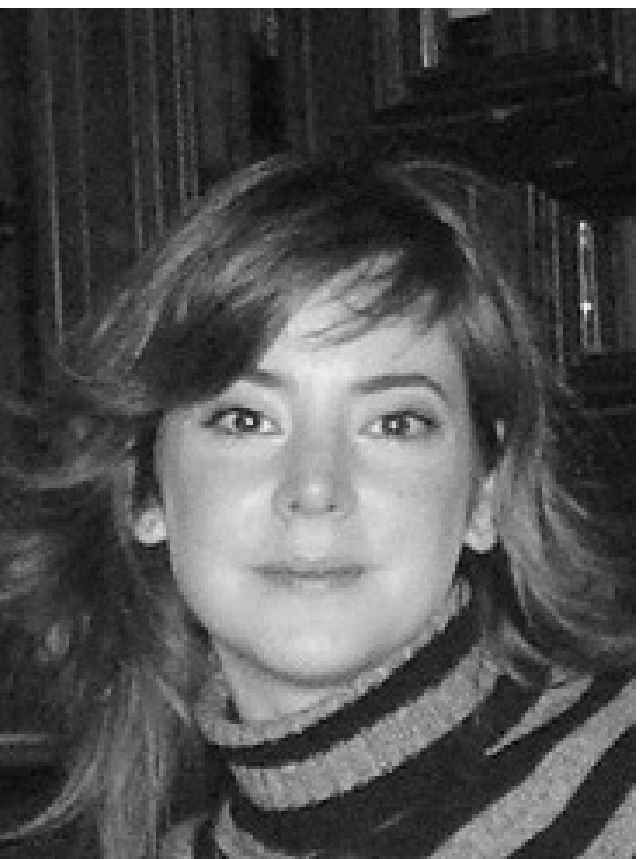}}]{Raquel Garc\'{\i}a-Bertrand}
(S'02--M'06--SM'12) received the Ingeniera Industrial degree and the Ph.D.
degree from the Universidad de Castilla-La Mancha, Ciudad Real,
Spain, in 2001 and 2005, respectively.

She is currently an Associate Professor of electrical engineering at
the Universidad de Castilla-La Mancha. Her research interests
include operations, planning, and economics of electric
energy systems, as well as optimization and decomposition
techniques.
\end{IEEEbiography}

\vspace{-1cm}

\begin{IEEEbiography}[{\includegraphics[width=1in,height=1.25in,clip,keepaspectratio]{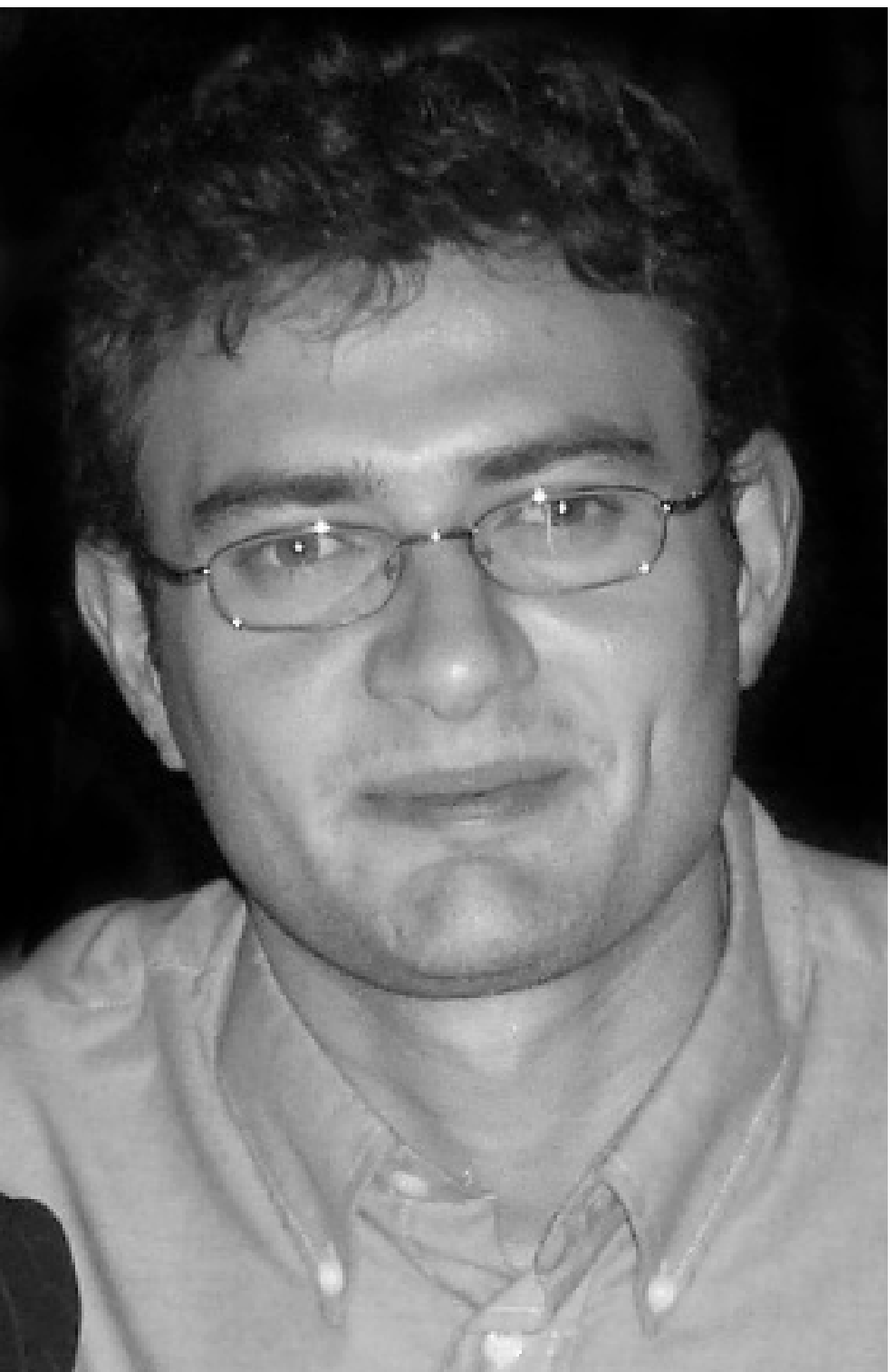}}]{Roberto M\'{\i}nguez} received the Civil Engineer degree and the Ph.D.
degree from the Universidad de Cantabria, Santander,
Spain, in 2000 and 2003, respectively.

He is currently a research fellow at the company Hidralab Ingeniería y Desarrollo, S.L., spin-off from the Universidad de Castilla-La Mancha. His research interests
include reliability engineering, sensitivity analysis, numerical methods, and optimization.
\end{IEEEbiography}

\end{document}